\documentclass[prb,twocolumn,aps,superscriptaddress,floatfix]{revtex4-2}

\usepackage{graphicx}
\usepackage{dcolumn}
\usepackage{bm}
\usepackage{amssymb}
\usepackage{amsmath}
\usepackage{subfigure}
\usepackage{physics}
\usepackage{sublabel}
\usepackage[utf8]{inputenc}
\usepackage{hyperref}
\usepackage[usenames, dvipsnames]{color}
\usepackage[english]{babel}
\usepackage[T1]{fontenc}
\usepackage{bbm}

\usepackage{verbatim}
\usepackage{xfrac}
\usepackage{upgreek}
\usepackage{float}

\hypersetup{colorlinks = true, linkcolor=blue, citecolor=blue, urlcolor=blue}

  \makeatletter
    \renewcommand\@make@capt@title[2]{%
     \@ifx@empty\float@link{\@firstofone}{\expandafter\href\expandafter{\float@link}}%
      {\textsc{#1}}\@caption@fignum@sep#2\quad}%
    \makeatother

\begin{document}


\title{The ``Gate of Fate'' - a quantum gaming concept }

\author{Deeptanshu Malu}
\author{Deevyanshu Malu}%
\affiliation{Department of Computer Science and Engineering, IIT Bombay, Powai, Mumbai-400076, India}
\author{Bhaskaran Muralidharan}
\affiliation{Department of Electrical Engineering and the Center of Excellence in Quantum Information Computing Science and Technology, IIT Bombay, Powai, Mumbai-400076, India}





\date{\today}

\begin{abstract}

Quantum physics underpins the emerging quantum technology era, yet its counterintuitive principles remain challenging for beginners. Video games offer a promising medium to make these principles accessible through an interactive experience. Conventional gameplay—even when featuring extraordinary abilities—largely draws on classical intuition, shaping the players' reflexes through familiar expectations of motion, action, and consequence. Here, we propose extending these reflexes into the quantum domain by embedding superposition, unitary evolution, quantum gates, and measurement into the game. Central to this approach is a new controller action, the “Gate of Fate,” which allows players to engage in quantum operations through their gaming decisions. We demonstrate the concept in a Tetris-based prototype, with progressively structured levels that introduce quantum operations and their consequences. We also outline extensions to role-playing games, in which the behavior and difficulty of antagonists can respond to players’ developing quantum gaming reflexes. By turning abstract quantum concepts into actionable game mechanics, this approach aims to cultivate quantum intuition through play, offering opportunities for both quantum pedagogy and more sophisticated game design.

\end{abstract}

\maketitle


\section{Introduction\label{sec:introdution}}

\begin{figure*}[htbp]
    \centering
    \includegraphics[width=0.9\linewidth]{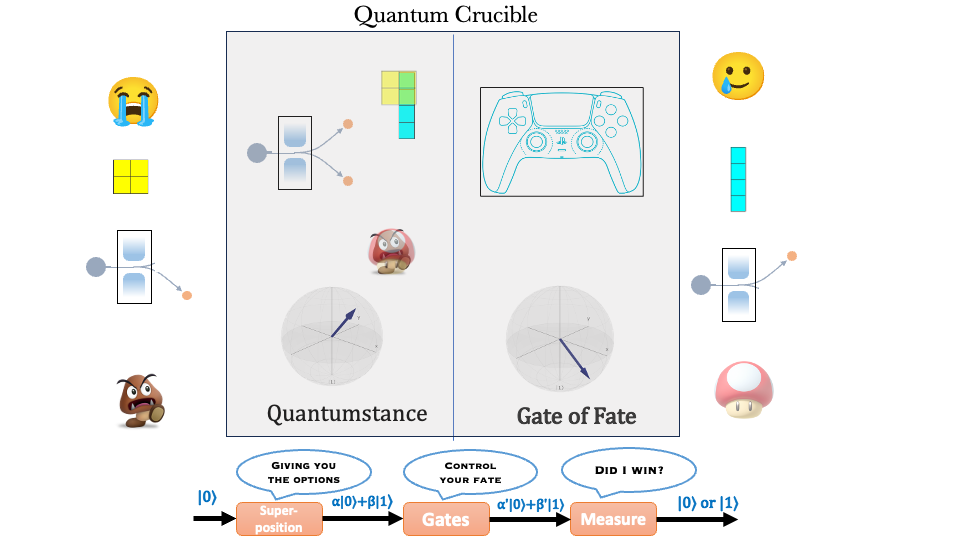}
    \caption{
   The quantum crucible comprises two key components: the Quantumstance and the Gate of Fate. This framework enables transforming an undesirable state into a more preferable one via quantum operations. An overview of the overall pipeline is presented below the crucible.
    }
    \label{fig:quantum_crucible}
\end{figure*}

Quantum superposition and interference offer possibilities for novel narratives and interactive experiences in video games, with potential extensions to cinema, musical tone generation, and perception. In conventional gaming, players typically respond to the outcomes of chance: they accept the roll of the dice and adapt their strategies accordingly. What if players could access an additional layer of control, shaping the possibilities from which an outcome emerges? Here, we introduce the “Quantum Crucible,” a framework for making quantum physics accessible through gameplay. By incorporating superposition, quantum operations, and measurement into game mechanics, the framework allows players to manipulate probability amplitudes, and thereby influence the likelihood of different outcomes. The aim is to transform chance from a condition players merely confront into a space they can actively navigate, cultivating quantum intuition through strategic play.  \\
\indent Conventional video gaming is shaped by two largely classical features. The first is how choices and challenges arise: encounters, adversaries, and events are often generated randomly, leaving players to respond to the outcomes presented. The second is how players exercise control: gameplay typically follows classically intuitive rules, and skilled players develop muscle memory to act swiftly within them. This venture introduces quantum principles into both features. Quantum operations allow players to influence the probabilities governing the choices and challenges they encounter, while new controller actions invite them to develop reflexes grounded in quantum mechanics. The aim is to make the “space of fate” itself playable: players learn not only to respond to unfolding events, but also to shape the possibilities from which those events emerge. \\
\indent To realize this idea, we introduce an additional quantum layer within a gaming setup, which we call the “quantum crucible”—a space in which players use quantum principles to influence their fate. Figure~\ref{fig:quantum_crucible} illustrates the concept from left to right. When a game presents an unfavorable outcome, the player enters the quantum crucible to find an opportunity to improve it. This process comprises three stages. First, favorable and unfavorable possibilities are encoded in a quantum superposition, establishing the “quantumstance.” Second, the player manipulates this superposition through quantum gate operations—the “Gate of Fate”—to reshape the probabilities of the available outcomes. Third, measurement produces a definitive outcome according to the resulting probabilities. Crucially, a skillful sequence of operations can increase the likelihood of a favorable result without necessarily guaranteeing it. The challenge is therefore to navigate the space of fate by learning how quantum operations shape the odds.\\
\indent We propose that the concepts developed here offer a distinctive direction for quantum-inspired gameplay and merit implementation. Appendix \ref{prior_art} summarizes the existing developments and situates our proposal within the larger landscape. Our literature and prior-art survey suggest that the combined use of quantum superposition and interference to shape gaming narratives, support quantum pedagogy, and explore quantum-based musical tone generation remains underexplored. Piispanen et al.~\cite{Piispanen_2025} identify three dimensions of quantum games: perceivable quantum physics, quantum technologies, and scientific purposes, including education and citizen science. Our concept of "Gate of Fate" aligns primarily with the third dimension through its pedagogical aim: helping players develop quantum intuition through interactive decisions and their consequences. A detailed prior-art search is available in App.~\ref{prior_art}.\\
\indent The remainder of this work develops these ideas systematically. The next section introduces the quantum crucible and the “Gate of Fate” framework. Section III presents a concrete implementation through our q-Tetris prototype. We then explore extensions that enrich gameplay through quantum principles. Finally, we outline broader possibilities for incorporating the “Gate of Fate” into diverse gaming settings.
\section{The Quantum Crucible}
We now describe the central proposal: using quantum physics to influence fate within a game. We introduce the \textit{Gate of Fate} (GOF), a mechanism through which players apply quantum gates to manipulate a superposition of game states, shaping the probabilities of outcomes obtained upon measurement. This mechanism forms part of the broader \textit{quantum crucible} framework illustrated in Fig.~\ref{fig:quantum_crucible}. During gameplay, a player may encounter an unfavorable situation, such as a Goomba in the Mario series or an inconvenient block in Tetris. Entering the quantum crucible offers an opportunity to increase the likelihood of a more favorable outcome. The framework comprises two key components: the \textit{quantumstance}, which establishes a superposition of possible game states, and the GOF, through which the player manipulates that superposition. A subsequent measurement determines the resulting state of the game.\\
\indent The \textit{quantumstance} takes an initial pure state representing an unfavorable game situation and transforms it into a superposition of favorable and unfavorable alternatives. The GOF then allows the player to manipulate the probability amplitudes of these alternatives through unitary evolution implemented by quantum gates. Finally, measurement selects an outcome according to the resulting probabilities, offering the possibility of obtaining a more favorable game state—for example, a Super Mushroom in the Super Mario series.

\begin{figure}[htbp]
    \centering
    \includegraphics[width=\linewidth]{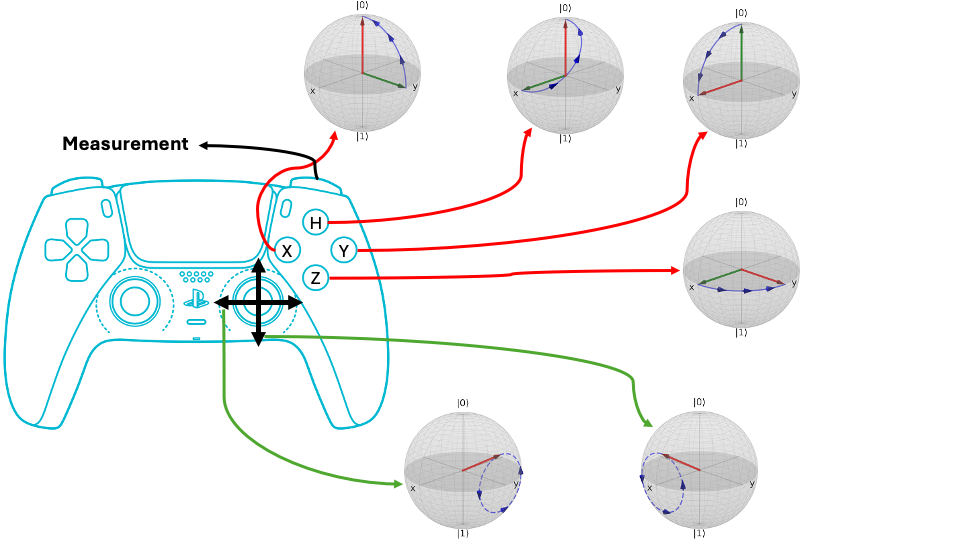}
    \caption{
    Mapping various buttons of a traditional game controller to their corresponding quantum operations, such as measurement and gate application.
    }
    \label{fig:quantum_controller}
\end{figure}

To enable the use of the GOF, we propose a mapping of various controls on a traditional game controller to various quantum operations as seen in Fig. ~\ref{fig:quantum_controller}. The Action buttons on the right are used to apply the quantum gates, namely $X$, $Y$, $Z$, and $H$ \cite{nielsen2010quantum} and the right joystick is used to perform unitary evolution about the x and y axes by moving it up and down and left-right, respectively. The top-right trigger is used to perform a measurement of the quantum superposition.\\
\indent Let us now introduce these concepts into the well known Tetris game, wherein the basic characters, the tetronimos serve as independent states and a superposition of whose can form states that can be manipulated by the user prior to the measurement stage. 
\hfill
\begin{figure*}[htbp]
    \centering
    \includegraphics[width=\linewidth]{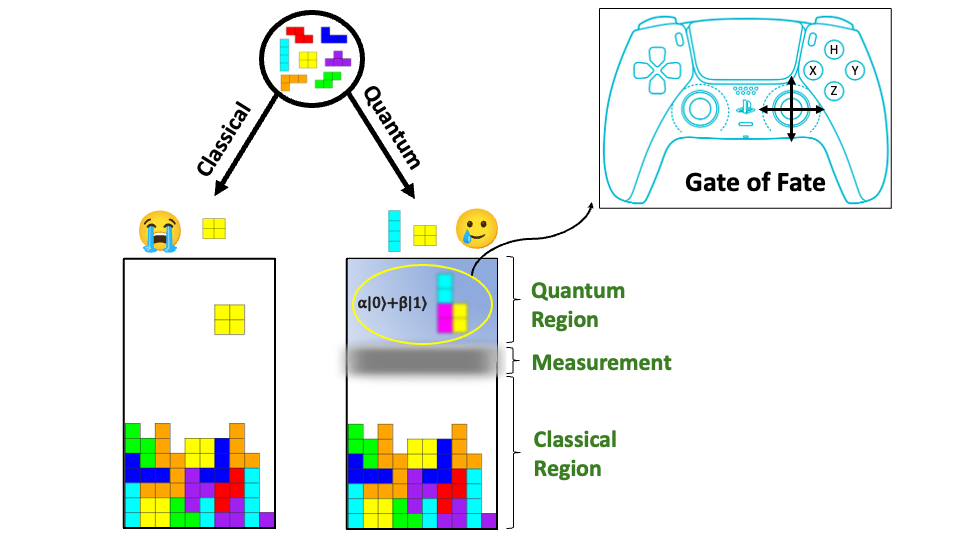}
    \caption{
    The quantum tetris: The main difference between Classical and Quantum Tetris. The three regions - the Quantum, Measurement, and Classical regions - that constitute the Quantum Tetris board are depicted. The GOF can only be applied in the Quantum region.
    }
    \label{fig:classical_vs_quantum_tetris}
\end{figure*}

\section{Quantum Tetris}
To demonstrate the concept of the GOF, we propose a quantum variant of the classical well-known game of Tetris, Quantum Tetris. The source code for our implementation was adapted from Tetromino, a Tetris clone developed by Al Sweigart~\cite{Tetromino}. A downloadable version of Quantum Tetris can be found at \url{https://prayog.swasth.org/qtetris}.\\
\indent In the classic game of Tetris, players manipulate one of seven kinds of falling tetrominoes by translating and rotating them to optimize their placement and maximize their score. But in Quantum Tetris, several core Tetris mechanics are reimagined through the lens of quantum computing. Key elements of the game, such as block generation, orientation, and selection, are governed by quantum principles, such as superposition, measurement, and unitary operations. \\
\indent In the original version of Tetris, a single unfavorable tetromino may be generated, while in Quantum Tetris, the superposition of an unfavorable and possibly a favorable tetromino may be generated, giving the player an opportunity to score higher. The player uses the GOF to manipulate the superposition to increase the chances of obtaining a favorable outcome. Figure \ref{fig:classical_vs_quantum_tetris} depicts the above phenomena. The game of Quantum Tetris consists of three levels of difficulty.

\subsection{Level 1}
In the first level of \emph{Quantum Tetris}, the generation of tetromino shapes, orientations, and colors is governed by quantum-inspired probabilistic circuits. Each circuit is initialized in an equal superposition state to ensure a uniform probability distribution over all admissible outcomes.\\
\indent The shape of the tetromino is determined using a 3-qubit circuit prepared in the state $\frac{1}{\sqrt{8}} \sum_{i=0}^{7} \ket{i}$. Upon measurement, one of the computational bases $\ket{000}$ to $\ket{111}$ is obtained. The seven non-zero outcomes, namely $\ket{001}$ through $\ket{111}$, are uniquely mapped to the seven standard tetromino shapes. If the result $\ket{000}$ is measured, it is discarded and the circuit is reinitialized and executed again. This rejection sampling procedure is repeated until a valid non-zero state is observed, thereby ensuring a uniform distribution across the seven tetromino shapes. Orientation and color are determined independently using separate 2-qubit circuits initialized to $\frac{1}{2} \sum_{i=0}^{3} \ket{i}$. Measurement yields one of the basis states $\ket{00}$ to $\ket{11}$, which are mapped to the four possible orientations and four possible colors (red, blue, green and yellow), respectively. \\
\indent Within the game area, two tetromino blocks are generated simultaneously and encoded into a single-qubit quantum state that represents their superposition. The computational basis states are defined such that $\ket{0}$ corresponds to the first tetromino and $\ket{1}$ corresponds to the second. The qubit is initialized to a normalized pure state $\ket{\psi} = \alpha \ket{0} + \beta \ket{1}$, where $\alpha, \beta \in \mathbb{C}$ satisfies $|\alpha|^2 + |\beta|^2 = 1$. The amplitudes $\alpha$ and $\beta$ are randomly selected subject to the normalization constraint, thus creating a probabilistic blend of the two tetrominoes. Visually, the overlapping cells of the superposed blocks are rendered in purple to indicate the presence of quantum superposition. This can be seen in the quantum region of Fig.~\ref{fig:classical_vs_quantum_tetris}. For each subsequent pair of tetrominoes, a new qubit is allocated to represent the next superposed state, and its amplitudes are determined only after the previous superposition has been measured.\\
\indent The gameplay environment is divided into a Quantum region and a Classical region. In the Quantum region, tetrominoes exist in a coherent superposition described by the single-qubit state. Players may apply unitary quantum operations to manipulate this state, specifically the rotation gates $R_x(\theta)$, $R_y(\theta)$, $R_z(\theta)$, and $H$ where $\theta \in \{90^\circ, -90^\circ\}$ and $H$ is the Hadamard gate. The applications of these gates with the help of a controller are visualized on Bloch spheres in Fig.~\ref{fig:quantum_controller}.\\
\indent A Bloch sphere is displayed to provide a geometric visualization of the qubit state, where any pure state can be expressed as $\ket{\psi} = \cos\left(\frac{\theta}{2}\right)\ket{0} + e^{i\phi}\sin\left(\frac{\theta}{2}\right)\ket{1}$. The application of rotation gates results in deterministic rotations about the corresponding axes, and the updated state is immediately reflected on the Bloch sphere to provide real-time visual feedback. \\
\indent When the superposed tetromino reaches the Classical region, a measurement is performed in the computational basis $\{\ket{0}, \ket{1}\}$. The state collapses to $\ket{0}$ with probability $|\alpha|^2$ or to $\ket{1}$ with probability $|\beta|^2$. Following measurement, only the resulting classical tetromino remains visible and interactable throughout the game. \\
\indent This integration of quantum state preparation, unitary evolution, and measurement-driven collapse not only influences gameplay dynamics but also provides an interactive pedagogical framework for illustrating fundamental principles of quantum mechanics.
\begin{figure*}[htbp]
    \centering
    \includegraphics[width=\linewidth]{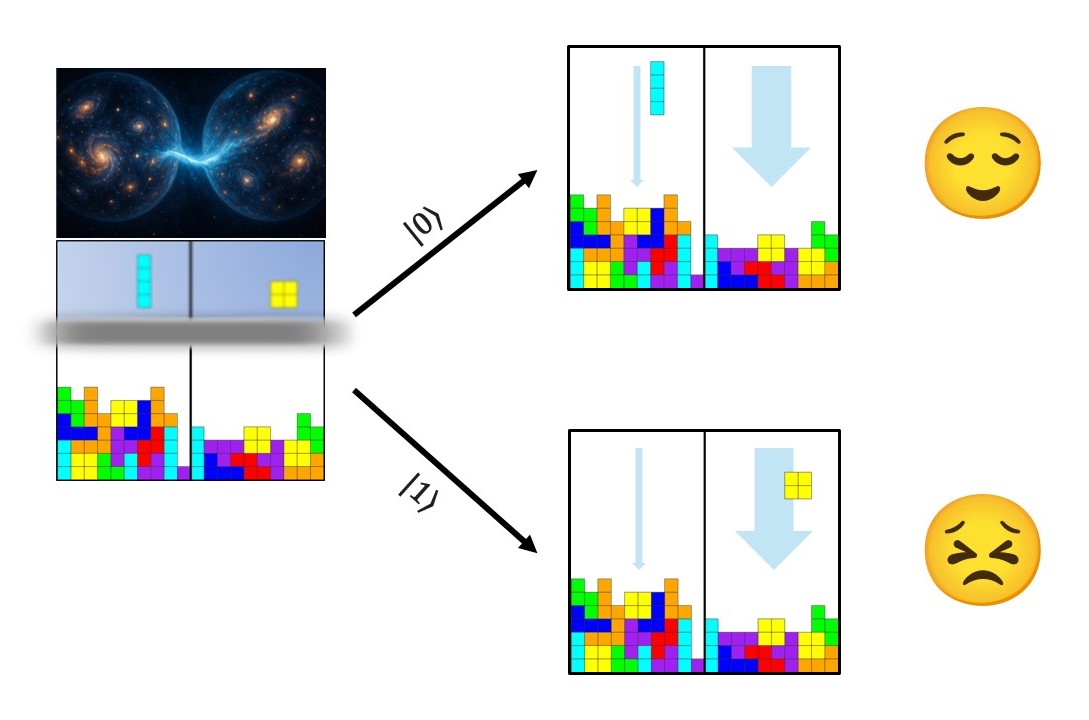}
    \caption{Depiction of the gameplay within a
"quantum multiverse", where the two
universes have different properties. In this
example, the universe on the right has an
adversarial setup with much higher gravity
than the one on the left.}
    \label{fig:quantum_multiverse}
\end{figure*}

\subsection{Level 2}

In this level, the concept of quantum evolution is introduced to illustrate how the state of a quantum system changes over time. Quantum evolution governs the time-dependent behavior of quantum states through unitary transformations. Conceptually, this may be viewed as analogous to the way gravitational forces influence the motion of classical objects, except that in quantum systems the evolution is described mathematically by unitary operators acting on the state vector.

In this level, the evolution of the quantum superposition is modeled by applying a small unitary transformation at each discrete time step. The state evolves according to $\ket{\psi'} = U\ket{\psi}$, where $U$ denotes the applied unitary operator. This evolution corresponds to a rotation of the state vector on the Bloch sphere representation. Specifically, the state undergoes a rotation described by the operator $R_X(1^\circ)$, i.e., a $1^\circ$ rotation about the X-axis, at every update of the game loop. This incremental rotation produces a continuous evolution of the qubit state on the Bloch sphere, gradually altering the probability amplitudes associated with the two tetromino blocks represented by $\ket{0}$ and $\ket{1}$. 

All other gameplay mechanics remain consistent with the previous level, including the Block Sphere visualization, the separation between the quantum and classical regions, and the procedures governing block structure and generation.

\subsection{Level 3}

In this level, the concept of Many-Worlds interpretation or the multiverse \cite{Gribbin2019SixImpossibleThings} is introduced into the gameplay. Two parallel game boards are instantiated, each representing a distinct universe. A single-qubit superposition of two tetrominoes is created in the same manner as in the previous level. However, in this setting each tetromino exists on a separate game board, with each board corresponding to one of the computational basis states of the qubit. Consequently, the tetromino associated with $\ket{0}$ resides in the first universe, while the tetromino associated with $\ket{1}$ resides in the second universe.

The division between the Quantum Region and the Classical Region remains present in this level. While the system is in the Quantum Region, the superposition persists and may be manipulated through the application of quantum gates, as described in the previous level. These unitary operations modify the probability amplitudes of the qubit state and therefore influence the likelihood of each universe being realized after measurement.

Once the superposed state enters the Classical Region, a projective measurement is performed in the computational basis. The quantum state collapses to either $\ket{0}$ or $\ket{1}$ with probabilities determined by the squared magnitudes of the corresponding amplitudes. Following measurement, the tetromino corresponding to the observed basis state materializes in its respective game board, while the alternative possibility is discarded.

The two boards then evolve independently, allowing the player to observe diverging gameplay trajectories arising from quantum measurement outcomes. This mechanism provides a visual representation of how different branches of a quantum system may lead to distinct classical realities.

Figure~\ref{fig:quantum_multiverse} illustrates two universes with distinct environments that are connected through the superposition of two tetromino blocks, one present in each universe. After measurement, the superposition collapses, leaving only the tetromino corresponding to the observed outcome in its respective universe. 

To introduce an additional gameplay challenge, the two universes operate with different block falling speeds. In one universe, the blocks descend more slowly, while in the other, they fall more rapidly. This asymmetry encourages the player to interpret one universe as relatively favorable and the other as unfavorable. Consequently, players are motivated to manipulate the quantum state using the available gates so as to increase the probability that the measurement outcome collapses to the tetromino associated with the more advantageous universe.

\subsection{Further level play and progression}

In each level, clearing a row of blocks awards the player one point, similar to the mechanics of classical Tetris. Whenever the player clears a predetermined number of lines, the falling speed of the tetrominoes increases, indicating progression through sub-levels. After completing a specified number of sub-levels, the player advances to the next level.

If the player is already in the final level, completing the required number of sub-levels results in victory. However, if at any point the stack of tetrominoes reaches the Classical--Quantum boundary, the player immediately loses and the game resets to Level 1. Consequently, to win the game, the player must successfully complete all three levels in a single uninterrupted playthrough.

\section{Extension of concepts to other game types- "Gate of Fate"}

Having expanded on the concepts of superposition, gates, and measurement in an abstract space, we now extend these possibilities into the spatial domain, specifically by taking up point-and-shoot play and role-playing games (RPGs).

\subsection{Spatial quantum evolution}

The concept of the GOF can be extended to projectile-based targeting games such as the Finnish throwing game Mölkky. In this setting, players typically aim to strike designated targets with a projectile, requiring both precision and strategic control.

In the second application illustrated in Fig.~\ref{fig:quantum_crucible}, we consider a simplified analogue of such a game. Here, a shooter emits a magnetic particle that traverses a region influenced by an external magnetic field, analogous to the Stern–Gerlach experiment. The objective is to guide the particle toward the upper (desirable) target. However, under the initial magnetic field configuration, the system exhibits a high probability of deflecting particles toward the lower (undesirable) target.

In this scenario, the GOF framework can be employed to systematically manipulate the magnetic field configuration. By changing the field parameters through the GOF, the probability distribution of particle trajectories can be reshaped, thereby increasing the likelihood of reaching the desired upper target.

\subsection{Role-playing games}
The concept of GOF can be illustrated through a game-based analogy inspired by the \textit{Mario} franchise, in which encounters with different entities lead to beneficial or adverse outcomes. A mushroom, for example, provides a power-up, whereas a Goomba represents an adversary. The third application in Fig.~\ref{fig:quantum_crucible} illustrates a simplified scenario in which the player initially encounters a Goomba, corresponding to an unfavourable outcome. Within the proposed GOF framework, this encounter is encoded in a superposed representation of favourable (mushroom) and unfavourable (Goomba) outcomes. Appropriate manipulation of this representation increases the probability associated with the favourable outcome. Upon measurement, the player is therefore more likely to obtain the mushroom and its associated benefit. This analogy illustrates how GOF can bias outcome probabilities toward desirable states, creating an opportunity to turn an initially disadvantageous encounter into a strategic advantage without guaranteeing a favourable result.
\section{Conclusion} To provide learners and technology enthusiasts with early exposure to quantum physics, we proposed integrating quantum concepts into video games, enabling players to develop gaming reflexes that incorporate superposition, unitary evolution, quantum gates, and measurement. We introduced a new controller action, the “Gate of Fate,” and illustrated its use through gameplay scenarios and potential extensions. Using Tetris as a testbed, we developed a simple prototype with a graded progression through levels that introduce different quantum operations. We also outlined extensions to role-playing games, where players’ “quantum gaming” reflexes could influence the hostile behaviour of antagonists. This approach offers a foundation for quantum pedagogy through interactive play, while opening avenues for more sophisticated quantum-inspired gaming experiences.
\section{Acknowledgements} \nonumber
The authors are grateful to S. Kapila for useful discussions and the creation of the user interface. The author BM acknowledges funding from the Department of Science and Technology (DST), Government of India, under the National Quantum Mission {through Grant no. DST/QTC / NQM/QMD/2024/4} and the Inani Chair Professorship fund {, through Grant No. DO/2024-INAN/001-001}.  The authors acknowledge funding from the Dhananjay Joshi Endowment award from IIT Bombay, {through Grant No: DO/2023-DJEF002}.
\section{Availability}
A downloadable version of Quantum Tetris, along with a user manual detailing the gameplay mechanics, is available online. Interested users can access and download these resources from the following link: \url{https://prayog.swasth.org/qtetris}.

\appendix
\section{Prior Art Search} \label{prior_art}
Quantum games have emerged over the past four decades as an interdisciplinary medium that blends play with the communication and exploration of quantum-mechanical principles. Early examples such as Atari’s Quantum (1982) and Quantum Soccer introduced wave functions and probabilistic motion through arcade-style mechanics \cite{Piispanen_2023}. This trajectory expanded to mobile and board games, including Universe Splitter, Quantum Race, qCraft, and quantum Tic-Tac-Toe, which employed metaphors of superposition, collapse, tunnelling, and entanglement to build player intuition rather than strict physical simulation \cite{chiarello2015board, goff2006}. \\
\indent Commercial and educational titles such as Quantum Break, Hello Quantum, Entanglion, and C.L.A.Y. further demonstrated how narrative, puzzles, and interactive challenges could be grounded in quantum concepts while remaining accessible to non-experts \cite{kamen, helloquantum, entanglion, wootton2020teach, clay2019, Piispanen_2025}. Together, these works establish quantum games as a pedagogical and cultural bridge between abstract theory and experiential learning. \\
\indent Beyond education, quantum games have also served as platforms for scientific participation and experimentation. Citizen-science projects such as Quantum Moves, Quantum Moves 2, meQuanics, Decodoku, and Bell’s Theorem collected large-scale human input to solve problems in quantum optimal control, error correction, and tests of nonlocality \cite{jensen2021, sherson2022, devitt2016, wootton2017gettingpublicinvolvedquantum, bellgame2018}. Command-line and puzzle-based games such as Quantum Battleship, Quantum Solitaire, and Cat/Box/Scissors explored entanglement, coherence, and quantum randomness through competitive or strategic gameplay \cite{wootton2017m, wootton2022}. \\
\indent Complementing these efforts, STAGE developed a suite of physical and digital games—Chicago Quant’em, 17, Quabble, Qunnect 4, Tailspin, Qubette, and the Quantum Photo Booth—that explicitly guided players through quantum key distribution and core notions of superposition and entanglement \cite{gaunkar2024gamedesigninspiredquantum, fuchs2020quantum, lin2020quantum, salimi2009investigation}. Collectively, prior art reveals recurring themes of using gameplay to build intuition, foster public engagement, and even generate data for real quantum research, highlighting the potential of game-based frameworks to both teach and probe quantum phenomena.\\
\bibliography{references}


\end{document}